# Superefficient electric-field-induced spin
# splitting in strained p-type quantum wells

D. M. Gvozdić and U. Ekenberg

Department of Microelectronics and Information Technology

Royal Institute of Technology

Electrum 229, SE-16440 Kista, Sweden

## Abstract

We investigate theoretically the efficiency of the Rashba effect, i.e. the spin splitting resulting from an electric field. This is the mechanism behind the Datta-Das spin transistor. In the research efforts so far the carriers have usually been taken to be electrons. Here we demonstrate remarkable improvements by several orders of magnitude by utilising holes instead. We also show that the frequently neglected lattice-mismatch between GaAs and AlGaAs can be used to further enhance the efficiency of the spin splitting mechanism. An inverse Rashba effect is also demonstrated.

In recent years there has been a strong increase of the interest in spin-related phenomena in solids. The prospects of combining semiconductor electronics with magnetic properties have lead to a new research area called "spintronics". Spin has been suggested to provide new functionalities for memories, electronic devices and quantum computation [1, 2]. It is well known that spin splitting occurs in the presence of an external magnetic field or magnetic atoms. It is also possible to create a conveniently controllable spin splitting with an electric field. This effect is of relativistic origin and is usually called the Rashba effect [3]. An applied electric field is seen in the frame of the moving carrier as a magnetic field. In a two-dimensional system the two-fold spin degeneracy then only remains when the in-plane wave vector $k$ is zero [4]. (See Fig. 1). One proposal that has lead to strong activities is the spin field effect transistor suggested by Datta and Das [5] which utilizes the Rashba effect. The efforts to implement this transistor in practice have not yet been very successful. An important task from both a fundamental and a technological point of view is to get a large spin splitting with a small electric field. The studies aiming at realizing spintronic devices have so far been concentrated to electron transport. In this Letter we show that using holes instead and introducing suitable strain we can obtain a remarkable enhancement of the efficiency of the Rashba effect by several orders of magnitude.

The spin transistor is similar to a normal field effect transistor but the carriers in the source and drain should be spin-polarized. Several mechanisms have been proposed to create this polarization [5-8]. There are presently strong activities to find suitable ferromagnetic semiconductor materials for source and drain [8-10]. A fundamental problem is that to get a large spin splitting it has so far been assumed that one needs semiconductor materials with large spin-orbit interaction (heavy atoms) while semiconductors with light atoms and magnetic ions (for source and drain) have been predicted to yield high Curie temperatures [10]. It would clearly be very valuable to get an efficient Rashba spin splitting in a material with moderately heavy atoms, e.g. GaAs.

The spin splitting of a subband can be viewed in two different ways, as shown in Fig. 1. So far people have usually considered the energy separation $\Delta E$ at a given $k$. However, the quantity that enters the theory of the spin transistor is the difference between the wave vectors, $\Delta k$, of a spin-split subband at a given energy [4, 5]. For electrons the spin splitting is frequently described by the Rashba term [3]

$$H_{so} = \alpha_{so}\, e\, \boldsymbol{\sigma} \cdot \mathbf{k} \times \boldsymbol{\varepsilon} \qquad (1)$$

where $\boldsymbol{\sigma}$ is given in terms of Pauli matrices, $e$ is the electron charge and $\alpha_{so}$ gives a measure of the strength of the spin-orbit interaction [11,12]. We take the uniform electric field $\boldsymbol{\varepsilon}$ to be perpendicular to the layers in a quantum well heterostructure. Adding Eq. (1) to a parabolic band with effective mass $m^*$ gives a spin splitting in $k$-space [5]

$$\Delta k = \frac{2\, m^*\, e\, \alpha_{so}\, \varepsilon}{\hbar^2} \qquad (2)$$

In this approximation $\Delta k$ is independent of energy, but it should be noted that this feature, that would be beneficial for the performance of spin transistors, is lost when nonparabolicity is taken into account. The thermal distribution of electrons at 300 K then gives a broadening of $\Delta k$ by about 5% around the mean value. The precession rate is proportional to $\Delta k$, and the precession length $L_\pi$ during which the desired reversal of the spin polarization takes place is given by $L_\pi = \pi/\Delta k$ [5].

We have performed subband calculations using an 8×8 $\mathbf{k}\cdot\mathbf{p}$ Hamiltonian which simultaneously yields an accurate description of the spin splitting of electron and hole subbands [13]. The interaction between the electron, heavy-hole, light-hole and split-off bands is included exactly in the matrix and the remote bands via perturbation theory. The potential of the quantum well and the electric field is added along the diagonal. We have applied the boundary conditions derived by Burt [14] and Foreman [15] and used a quadrature method [16]. The contribution to the spin splitting from the inversion asymmetry in the bulk crystal of III-V semiconductors [17] is negligible for holes in unstrained and weakly strained structures compared to the Rashba effect [18, 19].

Fig. 1 shows the spin-split energy dispersions for electrons (inset) and holes in a 20 nm GaAs quantum well. The interaction between heavy and light holes leads to strongly nonparabolic hole subband dispersions whose spin splitting clearly cannot be described by Eqs. (1) - (2). An analytical expression for heavy holes with $\Delta E$ proportional to $k^3$ [4] is displayed but it is found that the largest spin splitting occurs beyond its range of validity. The maximal $\Delta k$ is typically obtained a few meV below the top of the highest heavy-hole subband (HH1) which is also where the Fermi energy resides at common hole densities. Fig. 2 shows $\Delta k$ as a function of $\varepsilon$ for both electrons and holes. In agreement with the analytical expressions (1)-(2) we find that for electron subbands $\Delta k$ is almost independent of $E$ and strain and increases linearly with $\varepsilon$. The strength of the Rashba effect in a quantum well is not necessarily entirely given by bulk parameters. A previous approach to enhance the Rashba effect for electrons [20] utilized a small InAs quantum well asymmetrically positioned in a larger InGaAs/InAlAs quantum well. An enhancement of the Rashba parameter $\alpha$ [12] by a factor 3 compared to previous experiments was reported.

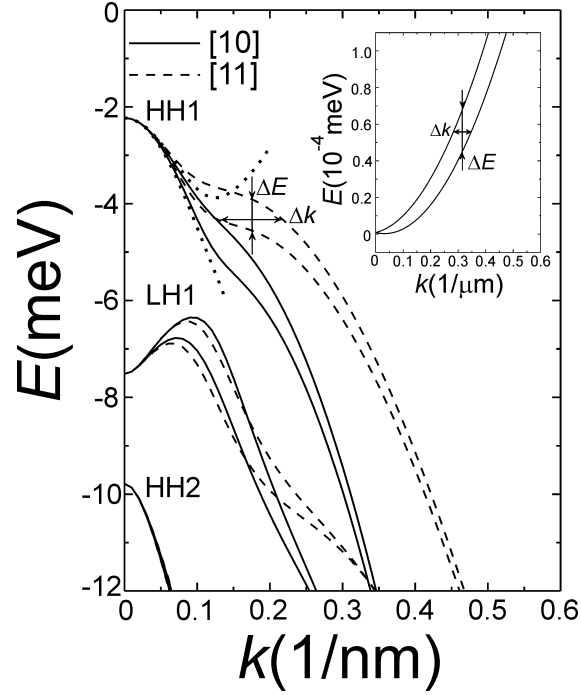

FIG. 1. Energy vs. in-plane wave vector for electrons (inset) and holes in a 20 nm wide unstrained GaAs quantum well between $Al_{0.22}Ga_{0.78}As$ barriers with an applied electric field of 5 kV/cm. Note the different scales for electrons and holes. Solid lines: [10] direction, dashed lines: [11] direction. The meanings of the spin splittings $\Delta E$ and $\Delta k$ are indicated. For holes we also display the effect of an analytical expression [4] where $\Delta E \propto k^3$ (dotted lines).

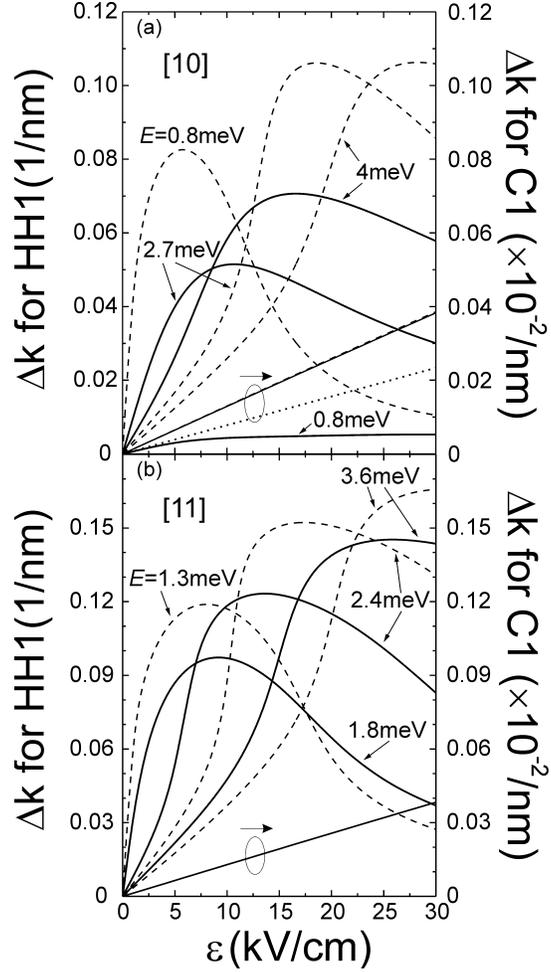

FIG. 2. Wave vector separation *Δk* vs. electric field along (a) the [10] direction and (b) the [11] direction in an unstrained (solid lines) and strained (dashed lines) GaAs quantum well. The hole energies relative to the top of the HH1 subband at which *Δk* is evaluated are shown. The corresponding results for electrons are essentially independent of energy. The dotted line for electrons in bulk GaAs is the analytical expression Eq. (2) with $\alpha_{so}$ taken from Ref. [11]. Note that the scales for electrons and holes differ by two orders of magnitude.

For holes we first discuss the case without strain (solid lines). *Δk* initially increases with electric field but for larger *ε* we find an inverse Rashba effect with *Δk* decreasing with *ε*. This phenomenon has recently been demonstrated experimentally [21]. A qualitative explanation is that the electric field increases the energy separation between HH1 and LH1 and thereby decreases the interaction between them. This mechanism can dominate over the influence of increasing inversion asymmetry.

The most important result in this work is that at a given electric field $\Delta E$ is much larger for holes than for electrons and $\Delta k$ is enhanced even more. One way to quantify this is to define an enhancement factor $\beta(\varepsilon)$ that tells how much stronger the splitting $\Delta k_h^{(max)}$ is for holes at the optimal energy compared to $\Delta k_e$ for electrons at the same $\varepsilon$. Using the analytical expression (2) we find

$$\beta(\varepsilon) = \frac{\Delta k_h^{(max)}(\varepsilon)}{\Delta k_e} = \frac{\hbar^2 \Delta k_h^{(max)}(\varepsilon)}{2 m^* e \alpha_{so} \varepsilon} = \frac{3.81 m_0}{m^* \alpha_{so}} \times \frac{\Delta k_h^{(max)}(\varepsilon)}{\varepsilon} \tag{3}$$

where the last step holds if we express $\alpha_{so}$ in Å$^2$. For GaAs $\alpha_{so} = 4.4$ Å$^2$ [11]. The stronger Rashba effect for holes compared to electrons has been noted by several authors [4, 20, 22, 23] but it has not been found previously that the enhancement can be so strong. Zawadski and Pfeffer [22] estimated that the Rashba spin splitting for holes would be enhanced over that of electrons by a factor $\sim$ (E$_0$ + $\Delta$)/$\Delta$, where E$_0$ is the band gap and $\Delta$ the spin-orbit splitting in the valence band. For GaAs this factor becomes 5.5 while we demonstrate further enhancement by up to three orders of magnitude. In Table I we present $\beta$ for some relevant cases. The main reason why this strong enhancement has not been noticed until now is that previously much stronger electric fields have been considered than where $\Delta k$ reaches its maximum. For a single modulation-doped interface with a typical areal hole density of 5·10$^{11}$ cm$^{-2}$ Poisson´s equation gives an electric field at the interface of 70 kV/cm. For a modulation-doped quantum well we can have a symmetric potential such that a small applied field can give the superefficient Rashba effect. Calculations for other materials have resulted in similar enhancements factors of 10$^2$ – 10$^3$.

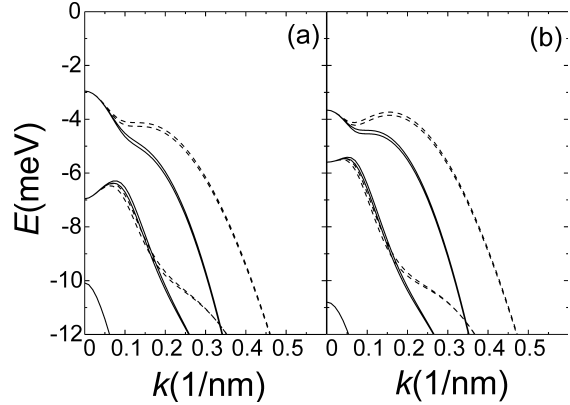

FIG. 3. Hole subband structure for a strained GaAs quantum well with $Al_{0.22}Ga_{0.78}As$ barriers in an electric field $\varepsilon$ of 1 kV/cm. In (a) we have an $Al_{0.07}Ga_{0.93}As$ buffer and a strain parameter $\zeta = 0.55$ meV such that flat dispersion is obtained along [11]. In (b) we have an $Al_{0.22}Ga_{0.78}As$ buffer, $\zeta = 1.6$ meV and flat dispersion along [10]. Solid lines: [10] direction, dashed lines: [11] direction.

For holes we demonstrate an even more efficient Rashba effect utilizing biaxial tension which moves LH1 and HH1 closer to each other. The increased repulsion makes it possible for the spin subbands of HH1 to have quite flat dispersions over a substantial range in $k$-space as is demonstrated in Fig. 3. $\Delta E$ increases gradually with electric field but with a flat dispersion even a small $\Delta E$ yields a substantial $\Delta k$ in this energy range. We have considered different directions in the two-dimensional Brillouin zone and found that the hole subbands are rather anisotropic. The largest $\Delta k$ is found along the [11] direction. In Fig. 3a the strain is small enough that the hole subband dispersions are monotonously decreasing with $k$ in all directions. In Fig. 3b we have chosen stronger biaxial tension in order to optimize the enhancement factor in the [10] direction. However, in this case the dispersion along the [11] direction is not monotonous and $\beta$ is not well-defined at all energies. It requires a closer analysis to determine if such subband dispersions can be useful for spintronic devices [24].

This strain-induced improvement is particularly clear and conveniently implemented in the well developed GaAs/Al$_x$Ga$_{1-x}$As system. The lattice-mismatch between GaAs and AlAs is only 0.2 % and usually neglected. These structures are usually grown on GaAs substrates such that the small biaxial compression of the Al$_x$Ga$_{1-x}$As barriers has negligible effect. But for growth on a sufficiently thick Al$_x$Ga$_{1-x}$As buffer layer, possibly on an InGaAs substrate with small In content, the lattice constant of the buffer material governs the in-plane lattice constant of the quantum well structure and the small strain in the GaAs layer gives surprisingly large effects, especially for the [10] direction, as shown by the dashed lines in Fig. 2. One advantage with this system is that a fine-adjustment of the lattice-mismatch becomes possible. With this structure as the active part of a spin transistor one can use the most developed ferromagnetic semiconductor GaMnAs as source and drain. Since Mn acts as an acceptor in AlGaAs this gives injection of holes, which is an additional advantage of p-type spin transistors. Successful hole injection in this system has been demonstrated [25].

In the optimal case in Table I ($\varepsilon = 2$ kV/cm, $E = 1.3$ meV, [11] direction) we find an enhancement factor $\beta = 5720$. The $\Delta k$-value corresponds to $L_\pi = 36$ nm which is comparable to the length of start-of-the-art conventional transistors [26]. $\beta$ tends to be larger for smaller $\varepsilon$. Another possible measure of the efficiency of the Rashba effect is $\partial(\Delta k)/\partial\varepsilon$ [27]. Replacing the last factor in Eq. (3) by this derivative one can easily reach $\beta > 10^4$.

There are also disadvantages of using holes. The main reason why n-type spin transistors have attracted much more attention than p-type ones is that the spin relaxation time for electrons is much longer than for holes [1, 2]. With the small $L_\pi$-value estimated above the transit time through a p-type spin transistor would only be of the order 1 ps. The hole spin relaxation time is known to be longer in heterostructures than in the bulk [28] and it also depends on doping conditions. Values between 4 ps and 1000 ps have been reported [28]. In a recent study of p-type quantum wells the conclusion was that it should be at least 100 ps [29].

Thus spin relaxation problems should not be insurmountable. The energy dependence of $\Delta k$ implies that the temperature sensitivity can be expected to become larger for holes than for electrons, resulting in a reduced on-off ratio of a p-type spin transistor in a similar way as incomplete spin injection, which has turned out to be a problem also for n-type devices. The larger hole masses can be expected to give worse transport properties but this drawback should be small compared to the much larger Rashba effect. For the short possible transistor length we also ought to take the quantization in the length direction into account instead of drawing conclusions from a two-dimensional subband structure.

In summary we have studied the efficiency of the Rashba effect for quantum wells with electrons and holes and found remarkable improvement by several orders of magnitude in the latter case. The small strain in GaAs/AlGaAs structures can be used to improve the results further. In spite of the remaining problems it seems clear that the much stronger Rashba effect for hole subbands calls for experimental and further theoretical efforts to examine fundamental spin properties and the feasibility of p-type spin transistors. For future work it could be worthwhile investigating more complex quantum well structures (e.g. asymmetric double quantum wells), quantum wires, structures grown along other directions than [001] etc.

We gratefully acknowledge financial support from the Swedish Research Council. We are grateful to P. Vogl for communicating results prior to publication and we also thank him, L. Thylén, B. Hessmo and U. Westergren for stimulating discussions.

<u>TABLE I</u>

Enhancement factor $\beta(\varepsilon)$ (Eq. (3)) for holes compared to electrons in a GaAs/Al$_{0.22}$Ga$_{0.78}$As quantum well for two electric fields and different quantum well structures. We display results in unstrained and strained quantum wells along two directions. The strain parameter $\zeta$ is defined such that the HH band edge is lowered by $\zeta$ meV relative to the unstrained case. ($\zeta > 0$ for biaxial tension).

| Direction | $\zeta$ (meV) | $\varepsilon = 2$ kV/cm | $\varepsilon = 5$ kV/cm |
|-----------|---------------|-------------------------|-------------------------|
| [10]      | 0             | 1237                    | 1036                    |
|           | 1.6           | 4175                    | 2128                    |
| [11]      | 0             | 3605                    | 2256                    |
|           | 0.55          | 5720                    | 2930                    |